\documentclass{moriond}
\usepackage{amsmath,amssymb,bm,multirow,float,bbold,hyperref}
\usepackage{graphicx}
\usepackage{xcolor,colortbl}
\definecolor{Gray}{gray}{0.95}
\definecolor{RGray}{gray}{0.85}
\definecolor{CGray}{gray}{0.92}

\def\be{\begin{equation}}
\def\ee{\end{equation}}
\def\bea{\begin{eqnarray}}
\def\eea{\end{eqnarray}}

\newcommand{\Photo}{}

\begin{document}
\vspace*{4cm}
\title{NON-UNIVERSAL $\mathbf{Z^\prime}$ MODELS WITH PROTECTED FLAVOUR-CHANGING INTERACTIONS}

\author{MARTIN JUNG}

\address{Excellence Cluster Universe, Technische Universit\"at M\"unchen,\\
  Boltzmannstr. 2, D-85748 Garching, Germany}

\maketitle\abstracts{
We define a new class of $Z^\prime$ models with neutral flavour-changing interactions at tree level in the down-quark sector. They are
related in an exact way to elements of the quark mixing matrix due to an underlying flavoured $U(1)^\prime$ gauge symmetry, rendering
these models particularly predictive. The same symmetry implies lepton-flavour non-universal couplings, fully determined by the gauge structure of
the model. Our models allow to address several presently observed deviations from the SM and specific correlations among the new
physics contributions to the Wilson coefficients $C^{(\prime)\ell}_{9,10}$ can be tested in $b\to s\ell^+\ell^-$ transitions. We furthermore
predict lepton-universality violations in $Z^\prime$ decays, testable at the LHC.
}

\section{Introduction}

The continuing success of the Standard Model (SM) 
has strong implications for new physics (NP) scenarios, which are consequently required to involve a very high mass scale, a highly
non-trivial flavour structure, or both. Two recent measruements by the LHCb collaboration involving semileptonic $b\to s\ell^+\ell^-$
transitions show deviations with respect to the SM expectations:
The ratio $R_K = \mathrm{Br}(B \rightarrow K \mu^+ \mu^-)/\mathrm{Br}(B\rightarrow K e^+ e^-)$ has been measured with a central value
indicating a violation of lepton universality at the $25\%$ level~\cite{Aaij:2014ora}, implying a $2.6\sigma$ deviation, and the
angular analysis of $B\to K^*\mu^+\mu^-$ decays, where specifically the observable $P'_5$ differs from the SM expectation with
$2.9\sigma$ significance~\cite{Aaij:2013qta}. Taking these measurements at face value, they require the NP to have (i) sizable
contributions to $b\to s\ell^+\ell^-$ transitions, and (ii) lepton non-universal couplings.

Extensions of the SM gauge group by an additional $\mathrm{U(1)}^\prime$ factor are among the possible NP scenarios that could explain
these
deviations. 
%
Such $\mathrm{U(1)}^\prime$ models have been popular extensions of the SM for many years~\cite{Langacker:2008yv}, but are in their
majority family universal, and therefore do not meet condition (ii). Models with
departures from family universality can give rise to large flavour-changing $Z^{\prime}$ interactions, which are strongly constrained
by current flavour data, as will be discussed below.
%
The conditions (i) and (ii) therefore limit considerably the viable $\mathrm{U(1)}^{\prime}$ gauge symmetries.

In order to avoid  exotic vector-like quarks, the $\mathrm{U(1)}^\prime$ symmetry has to involve both
quarks and leptons, and should be family-dependent in order to satisfy conditions (i) and (ii).
However, such a non-trivial flavour symmetry in the quark sector requires the extension of the scalar
sector in order to accommodate the quark masses and mixing angles~\cite{Leurer:1992wg,Felipe:2014zka}. It has been
shown in Ref.~\cite{Crivellin:2015lwa} that this option can be realized by including an additional
complex scalar doublet, yielding however up-quark FCNC which are not controlled by CKM elements or quark masses. 

We would like our model to have \emph{all} of its fermion couplings related \emph{exactly} to
elements of the CKM matrix. To that aim, we note that there is a class of two-Higgs-doublet models
(2HDM) that achieves this, known as Branco--Grimus--Lavoura (BGL) models~\cite{Branco:1996bq}. These
models have flavour-diagonal interactions in the up-quark and charged-lepton sectors, together with
flavour-changing neutral currents (FCNCs) in the down-quark sector. The latter are suppressed by
quark masses and/or off-diagonal CKM elements in an exact way, thereby providing an alternative
solution to the flavour problem in 2HDMs which differs radically from the hypothesis of natural
flavour conservation~\cite{Paschos:1976ay,Glashow:1976nt}.

While in the original BGL model the flavour symmetry is global, in this work we promote it to a local
one~\cite{Celis:2015ara}. This is achieved by charging also the leptons under the symmetry, thereby
enabling anomaly cancellation. In this gauged BGL framework ($\mathrm{U(1)}^\prime_{\rm BGL}$), the
properties of the BGL models are transferred to the gauge boson sector: we obtain FCNCs mediated at
tree-level by the neutral scalar and massive gauge vector bosons of the theory, all of which are
suppressed by off-diagonal CKM elements and/or fermion masses, and therefore naturally suppressed.
This class of models necessarily exhibits deviations from lepton universality due to its gauge
structure. 

\section{Gauged BGL symmetry}\label{s:frame}  
The characteristic features of BGL models~\cite{Branco:1996bq} are a consequence of specific patterns of Yukawa couplings, generated by
corresponding charge assignments under a horizontal, family-non-universal (BGL-)symmetry.
The quark Yukawa sector of the model reads ($i=1,2$)
$-\mathcal{L}^{\mbox{\scriptsize{quark}}}_{\mbox{\scriptsize{Yuk}}}=\overline{q_L^0}\, \Gamma_i\,
\Phi_i d^0_R+\overline{q_L^0}\, \Delta_i\, \widetilde{\Phi}_i  u^0_R+\text{h.c.} \,,$
where $\Gamma_i$ and $\Delta_i$ denote the Yukawa coupling matrices for the down- and up-quark sectors, respectively, and $\widetilde
\Phi_i \equiv i \sigma_2 \Phi_i^\ast$.
The neutral components of the Higgs doublets acquire vacuum expectation values (vevs) $|\langle
\Phi_i^0  \rangle  | = v_i/\sqrt{2}$, with $\tan \beta = v_2/v_1$ and $v \equiv (v_1^2 + v_2^2)^{1/2}\simeq 246$~GeV.

Choosing an abelian symmetry under which a field $\psi$ transforms as
$\psi\rightarrow e^{i\mathcal{X}^{\psi}}\psi\,,$
the most general quark-sector symmetry transformations yielding the required textures are of the form
\begin{equation}  \label{eq:system}
\mathcal{X}_L^q=\frac{1}{2}\left[\mathrm{diag}\left(X_{uR},X_{uR},X_{tR}\right)+X_{dR}\,\mathbb{1}\right]\,,\quad
\mathcal{X}_R^u=\mathrm{diag}\left(X_{uR},X_{uR},X_{tR}\right)\,,\quad
\mathcal{X}_R^d=X_{dR}\,\mathbb{1}\,,
\end{equation}
with $X_{uR}\neq X_{tR}$. 
The Higgs doublets transform as $\mathcal{X}^\Phi
= \frac{1}{2}\mathrm{diag}\,\left(X_{uR}-X_{dR},X_{tR}-X_{dR}\right)$.
There are several possible implementations of this symmetry, related by permutations in flavour
space and exchanging up- and down-quark sectors; here,
for definiteness, the top quark has been singled out, yielding the required patterns.
These textures give rise to FCNCs in the down-quark sector, 
which are however suppressed by quark masses and off-diagonal elements of the third CKM
row~\cite{Branco:1996bq}.
This choice implies a particularly strong suppression of flavour-changing phenomena in light-quark systems. Present
constraints from $\Delta F=1$ and $\Delta F=2$ quark flavour transitions are accommodated even when
the scalars of the theory are light, with masses of $\mathcal{O}(10^2)$~GeV~\cite{Botella:2014ska}.

When gauging a symmetry, special attention must be paid to whether it remains
anomaly-free. BGL models are automatically free of QCD anomalies, i.e.
$\mathrm{U(1)}^\prime[\mathrm{SU(3)}_C]^2$~\cite{Celis:2014iua}. However, the anomaly conditions for the following combinations need
to be fulfilled as well:
$\mathrm{U(1)}^\prime [\mathrm{SU(2)}_L]^2$, $\mathrm{U(1)}^\prime [\mathrm{U(1)}_Y]^2$,  
$[\mathrm{U(1)}^\prime]^2 \mathrm{U(1)}_Y $, $[\mathrm{U(1)}^\prime]^3$, $\mathrm{U(1)}^\prime [\mbox{Gravity}]^2$.
We find that there is no solution for this system within the quark sector alone  with the charge assignments in Eq.~\eqref{eq:system}.
Satisfying these anomaly conditions is highly non-trivial and requires, in general, additional fermions. However, the implementation of
the BGL symmetry as a local symmetry is possible by adding only the SM leptonic sector when allowing for lepton-flavour non-universal
couplings. 
Gauging the BGL symmetry exhibits the following features, beyond removing the unwanted Goldstone boson, which is ``eaten'' 
by the $Z^\prime$:
(i) A massive $Z^\prime$ with the SM fermion content, \emph{i.e.} no vector-like quarks are necessary.
(ii) Flavour-diagonal $Z^\prime$ couplings in the up-quark and charged-lepton sector, FCNC's in the down-quark sector, strongly
  suppressed by CKM matrix elements.
(iii) The gauge structure yields lepton-universality violation without the introduction of lepton-flavour violation, thereby
  providing an explicit counter-example to the general arguments given in 
  Refs.~\cite{Glashow:2014iga,Bhattacharya:2014wla,Boucenna:2015raa}.
(iv) All flavour couplings but one are fixed due to the gauge symmetry when including the anomaly conditions. The remaining one is
  fixed here via the condition $X_{\Phi_2}=0$ for definiteness. 
  The only remaining freedom lies in discrete lepton-flavour permutations of the $\mathrm{U(1)}^\prime$ charges,
  yielding six model implementations with identical quark and scalar charges.
In order to present the predictions for each of these models, we introduce generic lepton
charges $X_{i\,L,R}$,
such that each implementation is labeled by $(e,\,\mu,\,\tau)=(i,\,j,\,k)$. 
For numerical values and additional possibilities to implement the quark and neutrino sectors, see Ref.~\cite{Celis:2015ara}.

\section{Phenomenology} \label{sec:pheno}
The phenomenology of a scalar sector with a flavour structure identical to the one of our model has been analyzed in
Ref.~\cite{Botella:2014ska}. Since we are assuming a decoupling scalar sector, we can naturally accommodate a SM-like Higgs at
$125$~GeV; the heavy scalars are not expected to yield sizable contributions to flavour observables in general. A potential exception
are $B_{d,s}\to \mu^+\mu^-$  decays, as discussed in Ref.~\cite{Celis:2015ara}. We focus here on the phenomenological
implications of the $Z^{\prime}$ boson for this class of models.

Despite the large mass of several TeV, the $Z^\prime$ boson yields potentially significant contributions to flavour observables, due to
its flavour-violating couplings in the down-type quark sector. However, because of $M_W^2/M_{Z'}^2\leq 0.1\%$ and the CKM suppression of
the flavour-changing $Z'$ couplings, these contributions can only be relevant when the corresponding SM amplitude has a strong
suppression in addition to $G_F$. This is  specifically the case for meson mixing amplitudes and electroweak penguin processes which we
will discuss below. Further examples are differences of observables that are small in the SM, for example due to lepton universality or
isospin symmetry. On the other hand, we observe that the global CKM fit remains essentially valid in our models~\cite{Celis:2015ara}.
The particular flavour structure of the model implies a strong hierarchy for the size of different flavour
transitions, similar to the situation in the SM.

\begin{figure}
\parbox{8.1cm}{
\includegraphics[width=8.cm,height=8.cm]{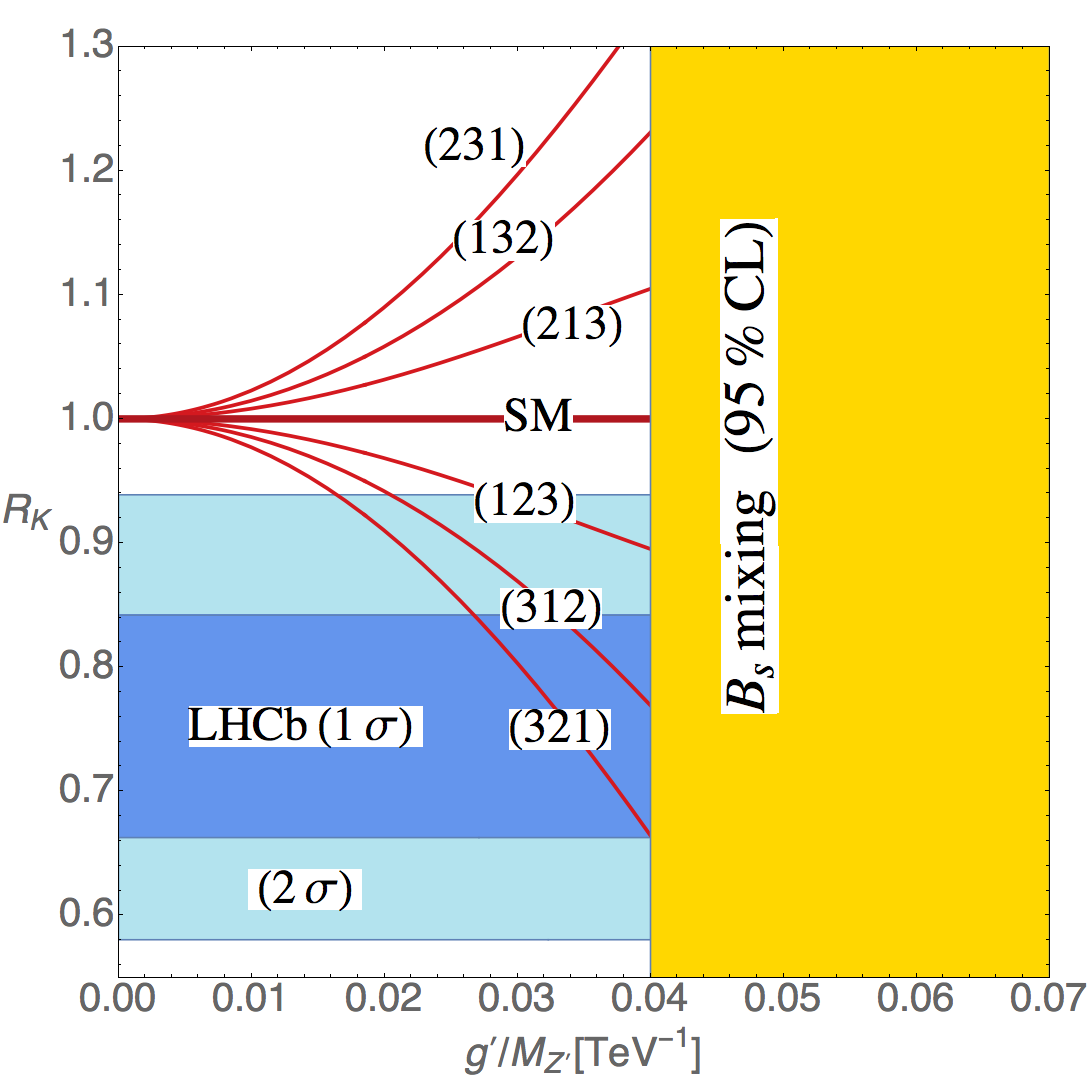}
}
\hspace{.5cm}
\parbox{7.cm}{
{\begin{tabular}{|c|c|c|c|c|c|} \rowcolor{RGray}
\hline 
Model   & $C_{9}^{\mbox{\scriptsize{NP}}\mu} (1\sigma)$  & $C_{9}^{\mbox{\scriptsize{NP}}\mu}  (2 \sigma)$     \\  \hline 
(1,2,3)   &  -- & $[-1.20, -0.61]$    \\[0,1cm]
(3,1,2) & $[-0.63, -0.43]$  &  $[-0.63, -0.17]$  \\[0.1cm] 
(3,2,1) & $[-1.20, -0.53]$  & $[-1.20, -0.20]$ \\
\hline
\end{tabular}
}\newline
\vspace{0.1cm}

{{\it \small Table~1: Model-dependent bound on $C_{9}^{\mbox{\rm{\scriptsize{NP}}}\mu}$
from the $R_K$ measurement.   The constraint from $B_s$ mixing is taken into account. }}

\vspace{1.cm}

{\caption{\it \small Model-dependent predictions for $R_{K}$ as a function of
$g^{\prime}/M_{Z^{\prime}}$. The recent measurement of $R_K$ by the LHCb collaboration is shown at
$1 \sigma$ and $2 \,\sigma$.   Constraints from $B_s$ mixing are also shown at $95\%$~CL.
\label{fig:RKmodels}}}
}
\end{figure}

Given the high experimental precision of $\Delta m_s$~\cite{Amhis:2014hma}, the strength of the
corresponding constraint depends completely on our capability to predict the SM value.\footnote{The same holds for $\Delta
m_d,\epsilon_K$, which yield similar bounds.} 
Using the recently obtained values from Ref.~\cite{Bazavov:2016nty} for the hadronic matrix elements, the obtained limit strengthens to
$M_{Z^\prime}/ g^{\prime}\gtrsim 25~\text{TeV}$ ($95\%$~CL)  compared to Ref.~\cite{Celis:2015ara}, see also
Fig.~\ref{fig:RKmodels}. In our models this bound is stronger than the ones from neutrino trident production and atomic parity
violation~\cite{Celis:2015ara}; the fact that these constraints are directly comparable stems from the high predictivity of our model.
Further potentially strong bounds stem from electric dipole moments (EDMs) and the anomalous magnetic moment of the muon. Both can
be shown to receive small effects in our models~\cite{Botella:2012ab,Jung:2013hka,Botella:2014ska,Celis:2015ara}.

The obvious way to search directly for a $Z^\prime$ is via  a resonance peak in the invariant-mass distribution of its decay products.
At the LHC this experimental analysis is usually performed by the ATLAS~\cite{Aad:2014cka} and CMS~\cite{Khachatryan:2014fba}
collaborations for $Z'$ production in the $s$-channel in a rather model-independent way, but assuming validity of the narrow-width
approximation (NWA), negligible contributions of interference with the SM~\cite{Accomando:2013sfa}, and flavour-universal
$Z^\prime$ couplings to quarks. While the assumptions for these expressions are not exactly fulfilled in our model, they are
applicable when neglecting the small contributions proportional to the off-diagonal CKM matrix elements and
taking into account that large couplings are typically excluded by flavour constraints~\cite{Celis:2015ara}.
Using these bounds together with the constraint from $R_K$, we find that a $Z^{\prime}$ boson is excluded in our models
for $M_{Z^{\prime}} \lesssim 3$-$4$~TeV, depending on the lepton charge assignments, while $g^{\prime}$ should be
$\mathcal{O}(10^{-1})$. 

An important feature of our models are sizable contributions to $C_{9,10}^{\ell, {\rm NP}}\sim (g^\prime/M_{Z^\prime})^2$. The relative
size of all these contributions is again fixed by the gauge symmetry. The  neutral-current semileptonic $b
\rightarrow s \ell^+ \ell^-$ transitions allow for testing these coefficients in global fits. Furthermore, they provide precise tests
of lepton universality when considering ratios of the type
$R_M \equiv   \frac{\mathrm{Br}(   \bar B \rightarrow \bar M \mu^+ \mu^- )}{\mathrm{Br}(   \bar B
\rightarrow \bar M e^+ e^- )}\stackrel{\rm SM}{=}1+\mathcal O(m_\mu^2/m_b^2)\,,$
with $M\in\{K, K^*, X_s, K_0(1430),\ldots\}$~\cite{Hiller:2003js},
where many sources of uncertainties cancel when integrating over identical phase-space regions.
Additional sensitivity to the dynamics of NP can be obtained via
double-ratios~\cite{Hiller:2014ula}, 
$\widehat R_{M} \equiv \frac{ R_M}{R_K} \,.$
Since $C_{9,10}^{\prime \ell} =0$ in our models we have
$\widehat R_{K^*} = \widehat  R_{X_s} = \widehat  R_{K_0(1430)} =1$, providing an important test of
the flavour structure of our models. 
Our models also give clean predictions for the rare decay modes $B \rightarrow \{K, K^*, X_s\}
\nu \bar \nu$~\cite{Buras:2012jb}: 
we obtain again a universal value for all ratios
$R_M^\nu=\mathrm{Br}(B\to M\bar\nu\nu)/\mathrm{Br}(B\to M\bar\nu\nu)|_{\rm SM}$, $\mathcal{O}(10\%)$ 
from unity for $g^{\prime} \sim 0.1$ and $M_{Z^{\prime}} \sim \mathcal{O}(\text{TeV})$, due to cancellations in the sum over
neutrinos.

If a $Z^{\prime}$ boson is discovered during the next runs of the LHC, ratios of its decays to different leptons can be used to
discriminate the models presented here~\cite{Celis:2015ara}.

Attributing the measurement of $R_K$ solely to NP  already excludes some of our models, since it requires sizable non-universal
contributions with a specific sign. The flavour structure in each of our models is fixed, so half of them cannot
accommodate $R_K< 1$.  
This is illustrated in
Fig.~\ref{fig:RKmodels}, where it is additionally seen that a large deviation from $R_K=1$, as
indicated by the present central value and $1\sigma$ interval, can actually only be explained in two
of the remaining models. This strong impact shows the importance of further measurements of $R_M$
ratios.

\begin{figure}
\includegraphics[width=5.1cm,height=5.1cm]{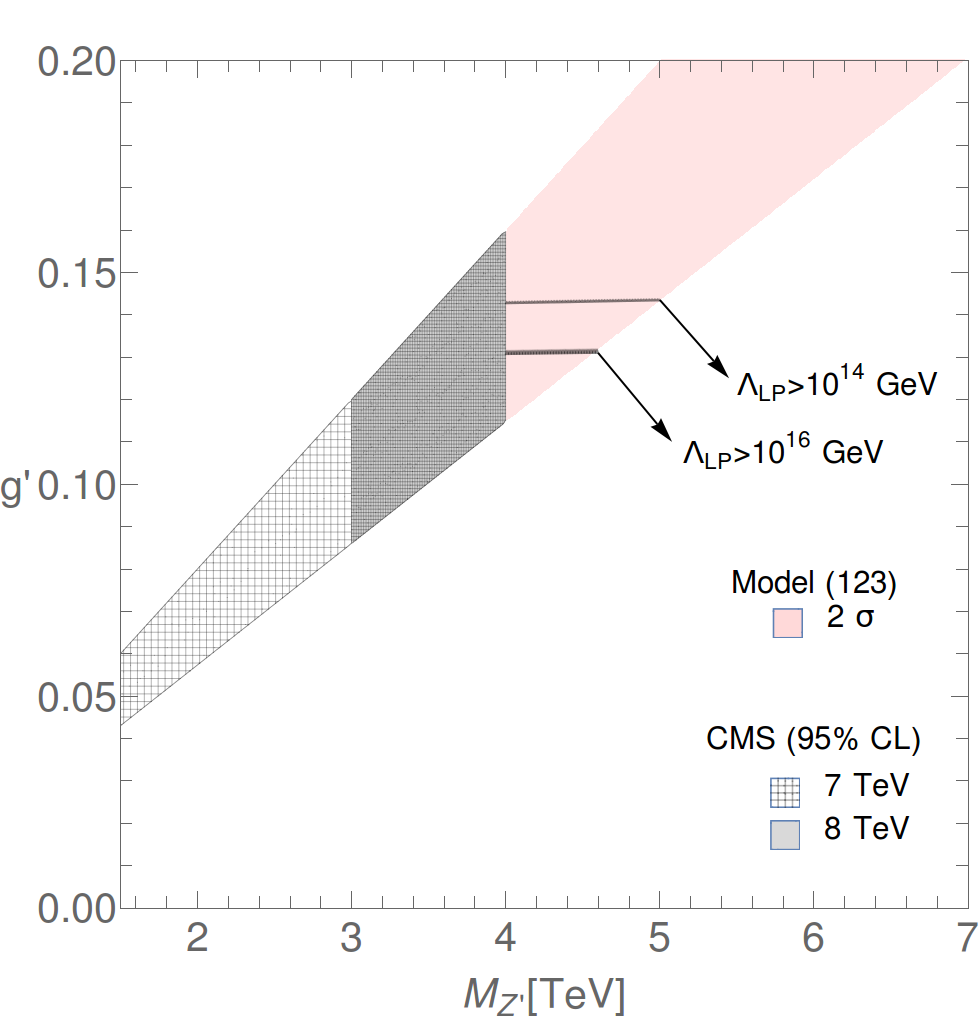}
~~%
\includegraphics[width=5.1cm,height=5.1cm]{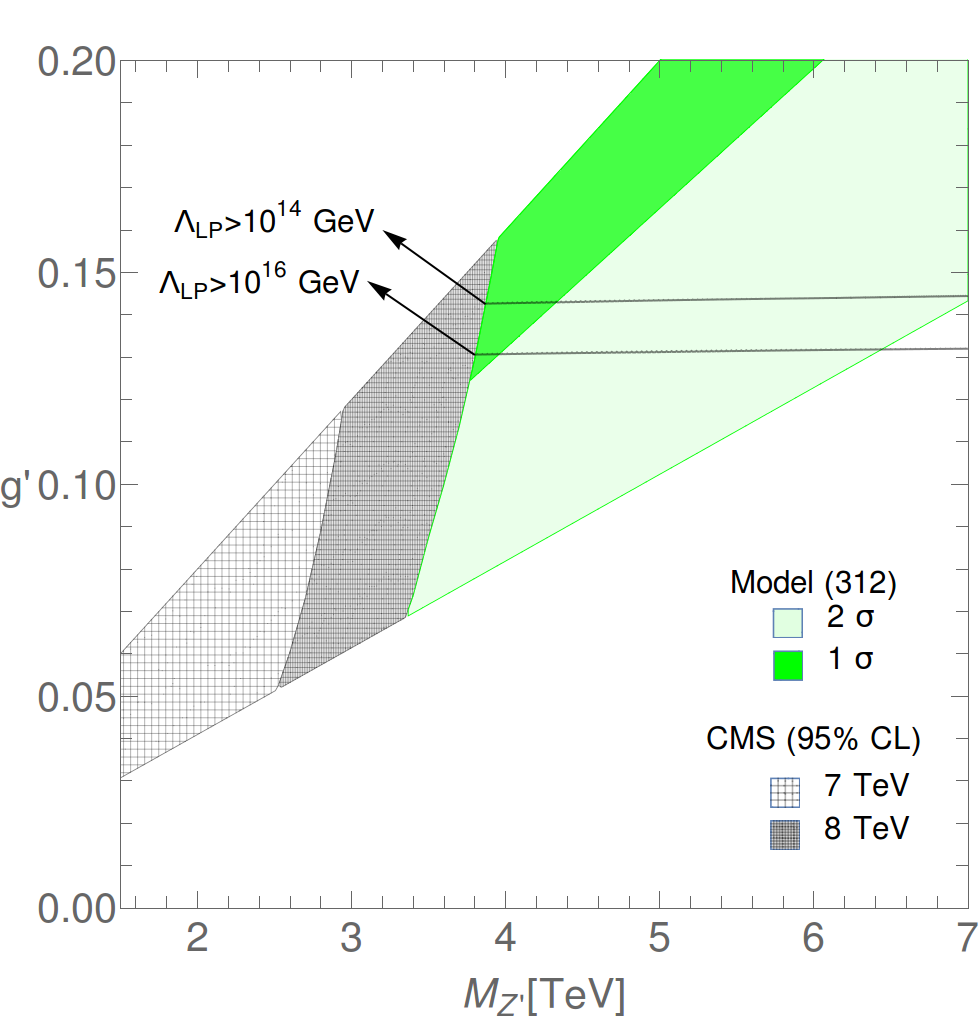}
~~%
\includegraphics[width=5.1cm,height=5.1cm]{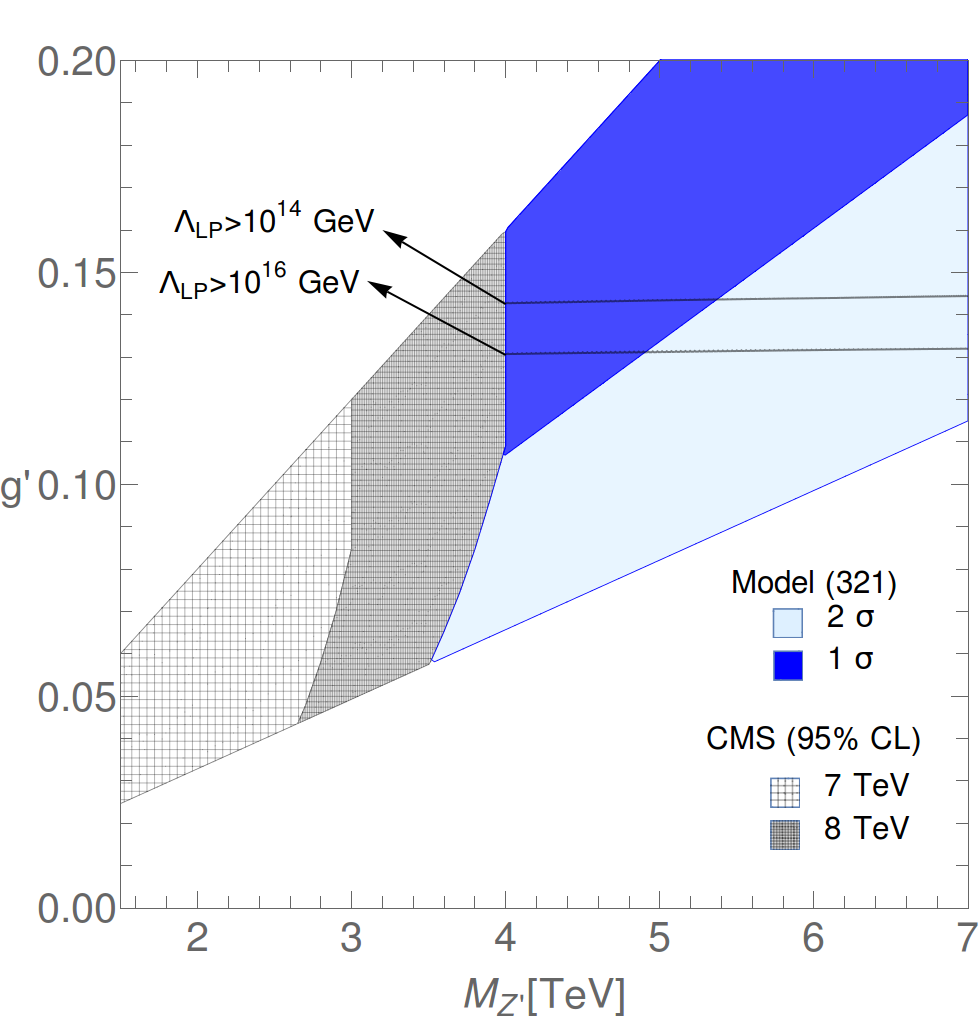}
\vspace{-3mm}
\caption{\it \small  Regions allowed at $1\sigma$ and $2\sigma$ by the $R_K$ measurement in the
$\{M_{Z^{\prime}},g^{\prime}\}$ plane for the models $(1,2,3)$, $(3,1,2)$ and $(3,2,1)$.   Exclusion
limits from $Z^{\prime}$ searches at the LHC are shown in gray. The black lines indicate bounds from
perturbativity of $g^\prime$.}  \label{fig:constraints}
\end{figure}

In Fig.~\ref{fig:constraints} we show the constraints from the $R_K$
measurement for the remaining models. The allowed
regions are consistent with the constraint from $B_s^0$-meson mixing. LHC searches
for a $Z^{\prime}$ boson exclude values of $M_{Z^{\prime}}$ below $3$-$4$~TeV, as discussed above; the corresponding areas are shown in
gray. We also show the theoretical perturbativity bounds obtained from the requirement that the Landau pole for the
$\mathrm{U(1)}^\prime$ gauge coupling appears beyond the see-saw or the Grand Unification scales, \emph{i.e.}
$\Lambda_{\mbox{\scriptsize{LP}}}>10^{14}$~GeV and $\Lambda_{\mbox{\scriptsize{LP}}}>10^{16}$~GeV, respectively.

Regarding the angular analysis in $B\to K^*\mu^+\mu^-$, we find that 
when  translating the $R_K$ measurement into a bound on $C_9^{{\rm NP}\mu}$ in our models, see Table~1, the
ranges are perfectly compatible with the values obtained from $P_5^\prime$, as also observed for other $Z^\prime$
models~\cite{Descotes-Genon:2013wba,Hurth:2013ssa,Altmannshofer:2015sma}.
This is highly non-trivial given the strong correlations in our models and will allow for decisive
tests with additional data.

\section{Conclusions}\label{sec:con}

The class of family-non-universal $Z^\prime$ models presented in this article exhibits FCNCs at tree
level that are in accordance with available flavour constraints while still inducing potentially
sizable effects in various processes, testable at existing and future colliders. This is achieved by
gauging the specific (BGL-)symmetry structure, introduced in Ref.~\cite{Branco:1996bq} for the first
time, which renders the resulting models highly predictive.
Additional phenomenological features are 
(i) non-universal lepton couplings determined by the charges under the additional $\mathrm{U(1)}^\prime$ symmetry, 
(ii) absence of FCNCs in the charged-lepton or up-quark sectors, and 
(iii) complete determination of the $\mathrm{U(1)}^\prime$ sector up to two real parameters, $M_{Z^\prime}$ and $g^\prime$, where all
observables at the electroweak scale and below depend only on the combination $g^\prime/M_{Z^\prime}$.

Present data already strongly restrict the possible parameter ranges in our models: direct searches
exclude $Z^\prime$ masses below $3-4$~TeV and the constraint from $B$ mixing implies
$M_{Z^\prime}/g^\prime\geq 25$~TeV ($95\%$ CL). 
Three of our models can explain the deviations from SM expectations in $b\to
s\ell^+\ell^-$ transitions seen in LHCb measurements \cite{Aaij:2013qta,Aaij:2014ora}, while the
other three are excluded (at $95\%$~CL) by $R_K<1$. 
These findings are illustrated in Figs.~\ref{fig:RKmodels} and~\ref{fig:constraints}. 

We predict in our models $ \widehat  R_M =1$, as well as universal ratios $R_M^\nu$ in $B\to M\bar \nu\nu$ decays, as well as specific
ratios for potential measurements of %
$\sigma(pp \rightarrow Z^{\prime} \rightarrow \ell_i \bar \ell_i )/\sigma(pp
\rightarrow Z^{\prime} \rightarrow \ell_j\bar \ell_j )$ at the LHC.
In the near future we will therefore be able to differentiate our new class of models from other
$Z^\prime$ models as well as its different realizations from each other. This will be possible due
to a combination of direct searches/measurements at the LHC and high-precision measurements at low
energies, \emph{e.g.} from Belle~II and LHCb. Further progress can come directly from theory,
\emph{e.g.} by more precise predictions for $\Delta m_{d,s}$ or $\epsilon_K$.

\section*{Acknowledgments}

I would like to thank the organizers for a fruitful workshop, and Alejandro Celis, Javier Fuentes-Mart\'in as well as Hugo Ser\^odio
for an enjoyable collaboration. This research was supported by the DFG cluster of excellence ``Origin and Structure of the Universe''.

\section*{References}
\bibliography{moriond_Jung}

\end{document}